\documentclass[twocolumn,showpacs,nofootinbib]{revtex4}
\usepackage{graphicx}
\usepackage{amsfonts}
\usepackage{amssymb}
\usepackage{amsmath}

\begin{document}

\title{The Structure of the Big Bang from
\\ Higher-Dimensional Embeddings}

\author{Sanjeev S. Seahra}
\email[Email: ]{ssseahra@uwaterloo.ca}
\author{Paul S. Wesson}
\email[Email: ]{wesson@astro.uwaterloo.ca}%
\homepage{http://astro.uwaterloo.ca/~wesson}

\affiliation{Department of Physics, University of Waterloo, \\
Waterloo, Ontario, N2L 3G1, Canada}

\date{\today}

\setlength\arraycolsep{2pt}
\newcommand*{\di}{\partial}
\newcommand*{\al}{\alpha}

\renewcommand{\textfraction}{0.15}
\renewcommand{\topfraction}{0.6}

\begin{abstract}

We give relations for the embedding of spatially-flat
Friedmann-Robertson-Walker cosmological models of Einstein's
theory in flat manifolds of the type used in Kaluza-Klein theory.
We present embedding diagrams that depict different 4D universes
as hypersurfaces in a higher dimensional flat manifold. The
morphology of the hypersurfaces is found to depend on the
equation of state of the matter.  The hypersurfaces possess a
line-like curvature singularity infinitesimally close to the $t =
0^+$ 3-surface, where $t$ is the time expired since the big bang.
The family of timelike comoving geodesics on any given
hypersurface is found to have a caustic on the singular line,
which we conclude is the 5D position of the point-like big bang.

\end{abstract}

\pacs{04.20.Jb, 11.10.kk, 98.80.Dr}

\maketitle

\section{Introduction}

In four-dimensional general relativity as formulated by Einstein,
the singularity theorems tell us that the big bang was a unique
birth event for a universe whose matter is of the conventional
sort and obeys the energy conditions \cite{Haw73,Wal84}.  The
early stages of the standard Friedmann-Robertson-Walker (FRW)
models may be modified by inflation, basically because then there
is a period when the energy conditions are violated
\cite{Lin90,Wes99}.  The associated process of mass generation
can either be described at the particle level by quantum field
theory or at the macroscopic level by classical theory
\cite{Lin79,Bro79,Gut81,Hen83,Bon60,Wes85}.  The early stages of
standard cosmology may also be modified by the effects of quantum
gravity and quantum tunneling \cite{Gib93,Vil82}.  In this
regard, the appropriate formalism in 4D for early times can be
investigated by considering the problem in 5D \cite{Dar00}. Also,
the status of standard cosmological models for late times can be
clarified by considering how they are embedded in 5D
\cite{Lid97}. There has recently been renewed interest in 5D
manifolds of the type used originally in Kaluza-Klein theory
\cite{Wes99,App87}. This is partly because of the resurrection of
Campbell's theorem, which ensures a perfect local embedding of a
4D Einstein space in a 5D Kaluza-Klein space
\cite{Cam26,Rip95,Rom96}; and partly because of the implications
of higher-dimensional approaches such as superstrings and
supergravity \cite{Wit81,West86,Gre87,Duf96} as well as string
and membrane theory \cite{You00,Maa00,Cha00}.  Recently, several
authors have revisited the problem of embedding 4D FRW metrics in
5D Minkowski space with an eye towards determining the
geometrical structure of the big bang \cite{Lyn89,Rin00}.  For
the standard spatially-flat ($k = 0$) 4D FRW models the
appropriate class of 5D models was given in a notable paper by
Ponce de Leon \cite{Pon88a}.  His models are separable in time
$t$, space $\sigma$, and an extra coordinate $\ell$, and reduce
to the FRW ones on $\ell$ = constant hypersurfaces (henceforth
denoted by $\Sigma_\ell$). Interesting work has been done on
those and related models, both from the astrophysical side
\cite{Wes92,Wes94,Wes95} and the mathematical side
\cite{McM94,Col95,Abo96}.  One of the Ponce de Leon 5D solutions
when unembedded is the 4D Einstein-de Sitter solution, which is
in tolerable agreement with observations; but surprisingly this
and other Ponce de Leon solutions have been shown by fast
algebraic computer package \cite{Lak95} to be flat in 5D.  This
may appear unusual, but recent work has made it clear that a flat
and apparently empty 5D manifold may contain a
curved, 4D manifold \cite{Wes99,Lid97,McM94,Col95,%
Abo96,Liu95,Wes00}. However, while much work has been done on the
5D Ponce de Leon solutions, no systematic account has been given
of how they embed the 4D big bang.

The present paper addresses this issue.  The plan of the paper is
as follows:  Section \ref{sec:Ponce} introduces the Ponce de Leon
class of one-parameter 5D cosmological solutions to the vacuum
Einstein field equations.  $\Sigma_\ell$ hypersurfaces are shown
to be geometrically identical to spatially flat ($k=0$) FRW
cosmological models.  The type of matter in these complementary
4D models is determined by the value of the parameter $\alpha$,
and includes both inflationary and non-inflationary cases.  The
5D manifold is then shown to be identical to Minkowski space
${\mathbb{M}}_5$ via a coordinate transformation. This coordinate
transformation allows us to plot our universe as a hypersurface
in a 3D Cartesian space once angular variables are suppressed,
which is the subject of section \ref{sec:structure}. Subsection
\ref{sec:singularities} deals with analytically determining where
comoving observers on $\Sigma_\ell$ emanate from in
${\mathbb{M}}_5$ as $t \rightarrow 0^+$, which corresponds to a
singularity in the 4-geometry. Subsection \ref{sec:geometry}
demonstrates that the local geometry of $\Sigma_\ell$ around
regular points is elliptical for non-inflationary cases and
hyperbolic for inflationary ones. Subsection \ref{sec:global}
establishes some results concerning the global properties of the
$\Sigma_\ell$ foliation of ${\mathbb{M}}_5$, the main conclusion
being that the cosmological coordinates used in the Ponce de Leon
metric do not cover all of 5D Minkowski space.  Subsection
\ref{sec:visualization} presents computer plots of $\Sigma_\ell$
hypersurfaces for a wide variety of cases.  These plots are
discussed in subsection \ref{sec:line}.  Also in this subsection,
a detailed analysis of the singular points on $\Sigma_\ell$ is
presented.  We find that there is a line-like curvature
singularity on $\Sigma_\ell$, but that the big bang corresponds
to only one point on that line. Section \ref{sec:special cases}
discusses special cases of the Ponce de Leon class not mentioned
in other sections and the corresponding 4D universes.  Section
\ref{sec:conclusion} contains our concluding remarks.

\paragraph*{Conventions.}
Throughout the paper, the 5D metric signature is taken to be
$(+----)$, while the choice of 4D metric signature is $(+---)$.
The spacetime coordinates are labeled $x^0 = t$, $x^{i} =
(r,\theta,\phi)$.  The extra coordinate is $x^4 = \ell$.  The
range of tensor indices is as follows: $A,B,\ldots = 0 - 4$;
$\alpha, \beta, \ldots = 0 - 3$; and $a,b = 0,1$.  $\nabla_A$
indicates the 5D covariant derivative.  We denote the metric of
flat 3-space by $d\sigma^2 = dr^2 + r^2 \, d\Omega^2$ with
$d\Omega^2 = d\theta^2 + \sin ^2 \theta \, d\phi^2$.  Finally, we
use geometric units where $c = G = 1$.

\section{Ponce de Leon Cosmologies}
\label{sec:Ponce}

The standard class of Ponce de Leon cosmological solutions is
flat in ordinary 3D space, curved in 4D spacetime and flat in 5D
\cite{Pon88a,Wes92,Wes94,Wes95,McM94,Col95,Abo96}.  It has a 5D
line element
\begin{equation}\label{5D Metric}
    dS^2 = \ell^2 \, dt^2 - t^{2/\alpha} \ell^{2/(1-\alpha)} \,
    d\sigma^2 - \alpha^2(1-\alpha)^{-2}t^2 \, d\ell^2.
\end{equation}
This metric is an exact solution of the 5D field equations in
apparent vacuum, which in terms of the Ricci tensor are
$R_{AB}=0$.  In what follows, we will examine the $\Sigma_\ell$
hypersurfaces of (\ref{5D Metric}) in detail.  The induced metric
$h_{\alpha\beta}$ on these hypersurfaces is given by
\begin{equation}\label{4D Metric}
    ds^2 = h_{\alpha\beta} dx^\alpha dx^\beta = \ell^2 dt^2 -
    t^{2/\al} \ell^{2/(1-\al)} d\sigma^2,
\end{equation}
which we can, of course, identify with the line element of a
spatially-flat ($k = 0$) FRW cosmology with scale factor
\begin{equation}
    a(\tau) = \ell^{(2\al - 1)/\al (1-\al)} \tau^{1/\al},
\end{equation}
where $\tau = \ell t$ corresponds to the FRW clock time.  That
is, the geometry of each of the $\Sigma_\ell$ 4-surfaces is
identical to geometry of the 4D spacetime models commonly used to
describe our universe.

We can use the form of the induced metric (\ref{4D Metric}) to
calculate the 4D Einstein tensor $G_{\alpha\beta}$ on each of the
$\Sigma_\ell$ surfaces. Alternatively, we can use standard
techniques to break the 5D field equations $R_{AB} = 0$ into one
wave equation, four conservation equations and ten equations for
the components of $G_{\alpha\beta}$ \cite{Wes99,Lid97,Cam26}. The
latter are given in general by
\begin{eqnarray}\nonumber
     G_{\alpha\beta} & = & \frac{\Phi_{;\alpha\beta}}{\Phi} +
     \frac{1}{2\Phi^2} \bigg\{ \frac{\Phi^* g^*_{\alpha\beta}}
     {\Phi} - \frac{ g^{\mu\nu} g^*_{\mu\nu}
     g^*_{\alpha\beta}}{2} +  g^{\lambda\mu} g^*_{\alpha\lambda}
     g^*_{\beta\mu} \\ & & - g^{**}_{\alpha\beta} + \frac{g_{\alpha\beta}}
     {4} \left[ g^{*\mu\nu} g^*_{\mu\nu} + \left( g^{\mu\nu} g^*_{\mu\nu}
     \right) ^2 \right] \bigg\} , \label{Induced Stress-Energy}
\end{eqnarray}
where $g_{44} = -\Phi^2$, and a star (${}^*$) denotes $\di / \di
\ell$. We now ask the question: what kind of 4D matter would give
rise to such an Einstein tensor?  To answer, we must associate
$G_{\alpha\beta}$ with a stress-energy tensor $T_{\alpha\beta}$
via the Einstein equations $G_{\alpha\beta} = 8\pi
T_{\alpha\beta}$.  For the Ponce de Leon solutions, this
``induced'' stress-energy tensor is of the perfect-fluid type
with the density and pressure of the ``induced'' matter given by:
\begin{equation}\label{Density and Pressure}
    8\pi\rho = \frac{3}{\alpha^2\tau^2}, \quad 8\pi p =
    \frac{2\alpha - 3}{\alpha^2\tau^2}.
\end{equation}
It should be stressed that this density and pressure refers to a
4D matter distribution that gives rise to a curved 4D manifold,
via the Einstein equation, geometrically identical to the
$\Sigma_\ell$ hypersurfaces.  We have not inserted any matter
into the 5D manifold by hand, as is commonly done in the
brane-world/thin-shell formalism where the 5D stress-energy is a
discontinuous, distribution-valued tensor field.  It is for this
reason that we call the matter described by $T_{\alpha\beta}$
induced; rather than being inserted into the Einstein equations
as an external source for the gravitational field, the matter
distribution is fixed by the morphology of the $\Sigma_\ell$
surfaces and the higher-dimensional geometry.

The equation of state of the induced matter is
\begin{equation}\label{EOS}
    p = (2\alpha/3 -1)\rho,
\end{equation}
which is determined by the dimensionless parameter $\alpha$ in
the metric (\ref{5D Metric}). For $\alpha = 3/2$, $8\pi\rho =
4/3\tau^2$ and $p=0$, which describes the late (dust) universe.
For $\alpha = 2$, $8\pi\rho = 3/4\tau^2$ and $8\pi p = 1/4\tau^2$
so $p = \rho/3$, which describes the early (radiation) universe.
That these quantities are the same as they are in the 4D FRW
models should not surprise us, as this could be inferred from
Campbell's theorem \cite{Wes99,Lid97,Cam26}.  However, the Ponce
de Leon class of solutions also embeds some more exotic
cosmological scenarios.  To see this, let us form the
gravitational density of the induced matter,
\begin{equation}\label{gravitational}
    \rho_g = \rho + 3p = \frac{3(\alpha-1)}{4\pi\alpha^2\tau^2}.
\end{equation}
The strong energy condition demands that $\rho_g \ge 0$, which is
only satisfied for $\alpha \ge 1$.  For cases which violate this
condition $\alpha \in (0,1)$, the Raychaudhuri equation says that
geodesics paths on $\Sigma_\ell$ accelerate away from one
another, as is expected in inflationary scenarios.  This is
easily verified by examining the congruence of comoving $d\sigma
= 0$ paths in the 4-geometry described by (\ref{4D Metric}). Just
like in conventional FRW cosmology, the proper distance between
adjacent paths will increase at an accelerating pace if the
deceleration parameter,
\begin{equation}\label{deceleration}
    q(\tau) \equiv - a \frac{d^2a}{d\tau^2} \left( \frac{da}{d\tau}
    \right)^{-2} = \alpha - 1,
\end{equation}
is negative.  We therefore conclude that models with $\alpha \in
(0,1)$ describe inflationary situations where the cosmological
fluid has repulsive properties, while models with $\alpha > 1$
have ordinary gravitating matter.  Note that we exclude models
with $\alpha < 0$ because they imply a contracting universe.

As a quick aside, we note that the Ponce de Leon metrics (\ref{5D
Metric}) have a constrained equation of state (i.e., the type of
matter corresponding to the curvature of the $\Sigma_\ell$
hypersurfaces does not change in time).  If one wanted to study a
more realistic model of our universe, the radiation-matter
transition could be generated by joining metrics with $\alpha =
2$ and $\alpha = 3/2$ across a $t =$ constant hypersurface that
represents the surface of last scattering. Such a calculation is,
however, beyond the scope of this paper.

It was discovered by computer that (\ref{5D Metric}) is actually
flat in 5D \cite{Wes92,Wes94,Wes95,McM94,Col95,Abo96,Lak95}.  To
prove this algebraically is non-trivial, but we can change
coordinates from those in (\ref{5D Metric}) to
\begin{subequations}\label{TL Transformation}
\begin{eqnarray}\nonumber
    T(t,r,\ell) & = & \frac{\al}{2} \left( 1 +
    \frac{r^2}{\al^2} \right) t^{1/\al} \ell^{1/(1-\al)} \\ \label{T} & - &
    \frac{\al}{2} \frac{ \left[ t^{-1} \ell^{\al/(1-\al)} \right]
    ^{(1-2\al)/\al}} {1 - 2\al}, \\ \label{R} R(t,r,\ell) & = & r
    t^{1/\al} \ell^{1/(1-\al)}, \\ \nonumber L(t,r,\ell) & = &
    \frac{\al}{2} \left( 1 - \frac{r^2}{\al^2} \right) t^{1/\al}
    \ell^{1/(1-\al)} \\ \label{L} & + & \frac{\al}{2} \frac{ \left[ t^{-1}
    \ell^{\al/(1-\al)} \right]^{(1-2\al)/\al}} {1 - 2\al}.
\end{eqnarray}
\end{subequations}
Then (\ref{5D Metric}) becomes
\begin{equation}\label{5D Minkowski}
    dS^2 = dT^2 - \left(dR^2 + R^2 \, d\Omega^2 \right) - dL^2,
\end{equation}
which is 5D Minkowski ${\mathbb{M}}_5$.  In order to contrast
with the standard coordinates $x^A = (t,r,\theta,\phi,\ell)$, we
will label the Minkowski coordinates as $y^A =
(T,R,\theta,\phi,L)$. The importance of (\ref{TL Transformation})
is that it allows us to visualize the geometric structure of our
universe when viewed from 5D flat space by plotting the surfaces
defined by $\ell =$ constant.  As seen in equation (\ref{4D
Metric}) above, these hypersurfaces share the same intrinsic
geometry as standard 4D FRW $k = 0$ cosmologies, which means that
they are equivalent to such models in a mathematical sense (this
is the paradigm commonly adopted when one tries to embed known
solutions of the Einstein equation in higher-dimensional
manifolds \cite{Lyn89,Rin00}).  We can therefore learn a great
deal about the local and global topological properties of 4D FRW
models, as well as the geometric structure of the ``physical''
big bang, from studying the $\Sigma_\ell$ hypersurfaces.  These
and other issues are considered in the following sections.

\section{The Structure of the Embedded Universes}
\label{sec:structure}

In order to analyze the geometrical properties of the
$\Sigma_\ell$ hypersurfaces, we note that the $T$, $R$, $L$
coordinates of (\ref{5D Minkowski}) are orthogonal and so can be
used as Cartesian axes. Then, the universe will have a shape
defined by $T = T(t,r,\ell_0)$, $R = R(t,r,\ell_0)$, $L =
L(t,r,\ell_0)$ on some hypersurface $\ell = \ell_0$.  The
following three subsections \ref{sec:singularities} --
\ref{sec:global} explore various characteristics of the
$\Sigma_\ell$ hypersurfaces analytically.  The goal is to get a
feel for the geometry of the $\Sigma_\ell$ hypersurfaces before
computer plots are presented in subsection
\ref{sec:visualization}.  In that subsection, we plot
hypersurfaces for several different cases, suppressing the
$\theta$ and $\phi$ coordinates so that our universe appears as a
2-surface embedded in Euclidean 3-space ${\mathbb{E}}_3$.
Subsection \ref{sec:line} deals with the location and
five-dimensional shape of the big-bang singularity, and is in
some sense a continuation of the discussion of singular points
started in subsection \ref{sec:singularities}.

\subsection{Properties of Singular Points on $\Sigma_\ell$}
\label{sec:singularities}

In this subsection, we will begin to investigate singularities on
the $\Sigma_\ell$ hypersurfaces, and introduce the problem of
locating where these singularities occur in ${\mathbb{M}}_5$. Our
goal is to get a feel for the shape of the $\Sigma_\ell$ surfaces
using analytical methods before resorting to the computer
plotting of subsection \ref{sec:visualization}.

A cursory inspection of equation (\ref{4D Metric}) reveals what
appears to be a singularity in the 4-geometry at $t = 0$, which
is commonly associated with the big bang.  We can calculate the
4D Kretschmann scalar (or curvature invariant) of the $\ell =
\ell_0$ hypersurfaces:
\begin{equation}\label{Kretschmann}
    K = R^{\alpha\beta\gamma\delta}R_{\alpha\beta\gamma\delta} =
    \frac{12 \left( \al^2 - 2\al + 2 \right)}{\ell_0^4\al^4t^4}.
\end{equation}
We note that $K$ diverges for all $\alpha$ at $t = 0$.  This
implies that the anomaly at $t = 0$ is a genuine curvature
singularity in the 4D manifold represented by (\ref{4D Metric}).
However, we know that the 5D manifold (\ref{5D Metric}) has
$R_{ABCD} = 0$ and is therefore devoid of any singularities.
Therefore, the 4D big bang results from the nature of the
$\Sigma_{\ell}$ hypersurfaces, or equivalently the choice of 5D
coordinates.

The question naturally arises of where the big bang at $t = 0$
appears in ${\mathbb{M}}_5$.  One way to approach the problem is
to take the $t \rightarrow 0^+$ limit of the coordinate
transformation (\ref{TL Transformation}) while holding $r$
constant.  It transpires that there are two different physically
interesting cases:
\begin{subequations}
\begin{eqnarray}\nonumber
    \lim_{t\rightarrow0^+} T(t,r,\ell) & = & - \frac{\alpha}{2(1-2\al)}
    \left[ \frac{\ell^{\al/(1-\al)}}{t} \right]^{(1-2\al)/\al} \\
    \label{T Limit} & \rightarrow &
    \begin{cases}
        0, & \alpha \in (\tfrac{1}{2},\infty) \\ -\infty, & \alpha \in
        (0,\tfrac{1}{2})
    \end{cases} \\ \label{R Limit}
    \lim_{t\rightarrow0^+} R(t,r,\ell) & = & 0, \quad \alpha > 0 \\
    \nonumber
    \lim_{t\rightarrow0^+} L(t,r,\ell) & = & + \frac{\alpha}{2(1-2\al)}
    \left[ \frac{\ell^{\al/(1-\al)}}{t} \right]^{(1-2\al)/\al} \\
    \label{L Limit} & \rightarrow &
    \begin{cases}
        0, & \alpha \in (\tfrac{1}{2},\infty) \\ +\infty, & \alpha
        \in (0,\tfrac{1}{2})
    \end{cases}.
\end{eqnarray}
\end{subequations}
In other words, the congruence of $(r,\theta,\phi) =$ constant
curves on $\Sigma_\ell$ (henceforth denoted by $\gamma_r$)
converges to a point as $t \rightarrow 0^+$ for all $\alpha > 0$.
For $\alpha \in (\tfrac{1}{2},\infty)$ the caustic is at $T = R =
L = 0$; for $\alpha \in (0,\tfrac{1}{2})$, the caustic is at null
infinity as approached by following the ray $T + L = R = 0$ into
the past. Na\"{\i}vely, this would seem to suggest that the the
big bang is a point-like event in 5D.  However, caution is
warranted because the limiting procedure described above is not
unique.  In particular, we need not approach $t=0$ in a manner
that leaves $r=$ constant.  For example, suppose we follow the
path $r(t) \propto t^{-1/2\al}$ into the past.  If $\alpha \in
(\tfrac{1}{2},\infty)$, then we will end up at the 5D point $T =
-L = \text{constant} \ne 0$, $R=0$, which is different from the
caustic in the $\gamma_r$ congruence. So, it is not correct to
uniquely associate $t = 0$ with the position of the $\gamma_r$
caustic.  In general, there are many points on $\Sigma_\ell$
eligible for that distinction. On the other hand, it should be
stressed that if we approach the initial singularity along any
path where $r(t)$ tends to a finite constant as $t \rightarrow
0^+$, then we will end up at one of the $\gamma_r$ caustics
(which one, depends on the value of $\alpha$). The issue of the
precise location of the big bang in 5D is complicated, so we will
defer a full discussion until subsection \ref{sec:line}.

Regardless of where the $t = 0$ surface lies in ${\mathbb{M}}_5$,
the calculations presented in this section demonstrate that the
point $T = R = L = 0$ is a special point for $\alpha \in
(\tfrac{1}{2},\infty)$, and that a point at null infinity $T = -L
\rightarrow - \infty, R = 0$ is a special point for $\alpha \in
(0,\tfrac{1}{2})$.  The computer-generated figures presented in
subsection \ref{sec:visualization} will bear out this analytical
conclusion.

\subsection{Properties of Regular Points on
$\Sigma_\ell$} \label{sec:geometry}

Having already given a partial discussion of the singular points
on the $\Sigma_\ell$ hypersurfaces, we would now like to
concentrate on analytically determining what $\Sigma_\ell$ looks
like in the neighborhood of regular points.  As mentioned above,
our plots of $\Sigma_\ell$ will look like 2-surfaces embedded in
${\mathbb{E}}_3$.  It is therefore useful to recall the notion of
the Gaussian curvature $G^{\mathbb{E}}$ of 2-surfaces $S \in
{\mathbb{E}}_3$ from standard differential geometry \cite{Lip69}.
Consider a point $P$ on a surface $S$ embedded in
${\mathbb{E}}_3$. In a small neighborhood around $P$, the surface
$S$ can be modeled as a paraboloid centered at $P$ (the so-called
``osculating paraboloid'').  The shape of this paraboloid is
given by the sign of the Gaussian curvature, which in turn is
given by the product of the principle curvatures of $S$ at $P$.
If $G^{\mathbb{E}}>0$, the paraboloid is elliptical and the
surface has a convex or concave shape about $P$.  If
$G^{\mathbb{E}}<0$, the paraboloid is hyperbolic and the local
shape of $S$ is that of the familiar ``saddle-surface''. If
$G^{\mathbb{E}}=0$, one or both of the principle curvatures are
zero implying a cylindrical or planar shape.  In terms of
intrinsic geometrical quantities, the Gaussian curvature is given
by
\begin{equation}\label{Gaussian}
    G^{\mathbb{E}} = \frac{{}^{(2)}R^{\mathbb{E}}_{0101}}
    {\det[{}^{(2)}g^{\mathbb{E}}_{ab}]},
\end{equation}
where ${}^{(2)}g^{\mathbb{E}}_{ab}$ is the Euclidean 2-metric on
$S$ and ${}^{(2)}R^{\mathbb{E}}_{1212}$ is the single independent
component of the associated 2D Riemann-Christoffel tensor. For
the Euclidean version of the $\Sigma_\ell$ hypersurfaces, we can
find ${}^{(2)}g^{\mathbb{E}}_{ab}$ from the positive-definite
line element
\begin{equation}
    {}^{(2)}g^{\mathbb{E}}_{ab} dx^a dx^b =
    \left( dT^2 + dR^2 + dL^2 \right)\Big|_{d\ell = 0}.
\end{equation}
Using this, we can calculate the Gaussian curvature
(\ref{Gaussian}) of the $\Sigma_\ell$ surfaces when embedded in
Euclidean space:
\begin{subequations}
\begin{equation}
    G^{\mathbb{E}} = \frac{4\al^6(\al-1) t^{2(\al-2)/\al}
    \ell^{-2(1+\al)/(1-\al)}} {\left[\chi^2 - 2 r^2\chi +
    (r^2 + \al^2)^2 \right]^2},
\end{equation}
where
\begin{equation}
    \chi \equiv \alpha^2 t^{2(\al - 1)/\al} \ell^{2\al/(\al-1)}.
\end{equation}
\end{subequations}
We note that the denominator of $G^{\mathbb{E}}$ is positive
definite for all values of $t$, $r$, $\ell$ and $\alpha$.
Therefore, $G^{\mathbb{E}} > 0$ for $\alpha \in (1,\infty)$ and
$G^{\mathbb{E}}<0$ for $\alpha \in (0,1)$ for all $(t,\ell)>0$.
In other words, when viewed in ${\mathbb{E}}_3$, all the
non-singular points on $\Sigma_\ell$ are elliptical for
non-inflationary models and hyperbolic for inflationary ones.

We mention in passing that it is also possible to calculate the
Gaussian curvature $G^{\mathbb{M}}$ of the $\Sigma_\ell$ surfaces
when they are embedded in ${\mathbb{M}}_3$.  The 2-metric
${}^{(2)} g_{ab}^{\mathbb{M}}$ is obtained by setting $d\theta =
d\phi = d\ell = 0$ in (\ref{5D Metric}).  This yields
\begin{equation}\label{Minkowski Gaussian}
    G^{\mathbb{M}} = \frac{{}^{(2)}R^{\mathbb{M}}_{0101}}
    {\det[{}^{(2)}g^{\mathbb{M}}_{ab}]} = \frac{\alpha-1}
    {\alpha^2t^2\ell^2},
\end{equation}
which, like $G^{\mathbb{E}}$, is positive for $\alpha \in
(1,\infty)$ and negative for $\alpha \in (0,1)$.  Also, we note
that $G^{\mathbb{M}}$ diverges as $t\rightarrow 0^+$ for all
$\alpha$, which is consistent with the behaviour of the 4D
Kretschmann scalar (\ref{Kretschmann}) calculated above.  The
Ricci tensor in 2D is given by ${}^{(2)}R_{ab}^{\mathbb{M}}
={}^{(2)} g_{ab}^{\mathbb{M}} G^{\mathbb{M}}$, so that the strong
energy condition reads ${}^{(2)}R_{ab}^{\mathbb{M}} u^a u^b =
G^{\mathbb{M}} \ge 0$ where $u^a$ is some timelike 2-velocity.
Hence, if $G^{\mathbb{M}} < 0$ then the strong energy condition
is violated in 2D. This implies that the 2D gravitational density
of the matter is less than zero, which in turn means that the
relative velocity between neighboring timelike geodesics is
increasing in magnitude by Raychaudhuri's equation. This is
consistent with the finding that $q(t) < 0$ and $\rho + 3p < 0$
for $\alpha \in (0,1)$ discussed in subsection \ref{sec:Ponce}.

In conclusion, we have found out that the local geometry around
regular points on $\Sigma_\ell$ -- for a given value of $\alpha$
-- is determined by whether or not the model represents a
universe with ordinary or inflationary matter.  In the former
case the regular points are elliptical, in the latter case they
are hyperbolic.

\subsection{Global Properties of $\Sigma_\ell$}
\label{sec:global}

In the previous two subsections, we have discussed the properties
of singular and regular points on $\Sigma_\ell$.  We now turn our
attention to the global properties of the hypersurfaces, with an
eye to determining which portion of ${\mathbb{M}}_5$ they
inhabit. In other words, we want to know what part of the
${\mathbb{M}}_5$ manifold is covered by the $x^A =
(t,r,\theta,\phi,\ell)$ coordinates.

The coordinate transformation (\ref{TL Transformation}) implies
\begin{equation}\label{T plus L}
(T+L)/R = \alpha/r.
\end{equation}
Since $R/r$ is positive for $(t,\ell) > 0$ and we assume that
$\alpha
> 0$, this implies that the $\Sigma_\ell$ hypersurfaces are
restricted to the half of the manifold defined by $(T+L)>0$.
Next, the transformation also implies that
\begin{equation}\label{light cone}
    T^2 - R^2 - L^2 = \frac{\alpha^2 t^2 \ell^2}{2\alpha - 1}.
\end{equation}
Hence, if $\alpha \in (\tfrac{1}{2},\infty)$ then the
$\Sigma_\ell$ surfaces must lie in the region satisfying $T^2 -
R^2 - L^2>0$, which is the volume inside the light cone
originating from $T=R=L=0$.  If $\alpha \in (0,\tfrac{1}{2})$,
the surfaces must lie in the region outside the light cone,
defined by $T^2 - R^2 - L^2<0$. Also, the $\Sigma_\ell$ surfaces
approach $T^2 - R^2 -L^2 = 0$ as $\ell \rightarrow 0$, implying
that the direction of increasing $\ell$ is always away from the
light cone.  Putting these facts together, we see that if $\alpha
\in (\tfrac{1}{2},\infty)$, then the coordinates utilized in
(\ref{5D Metric}) only cover the portion of ${\mathbb{M}}_5$
inside the light cone centered at the origin with $T>0$.  If
$\alpha \in (0,\tfrac{1}{2})$, then the coordinates in (\ref{5D
Metric}) cover the region of ${\mathbb{M}}_5$ satisfying $(T + L)
> 0$ and exterior to the light cone centered at the origin.
Finally, equations (\ref{R}), (\ref{T plus L}) and (\ref{light
cone}) can be rearranged to yield
\begin{equation}\label{holy grail}
    0 = UV - \frac{\alpha^{2(1-\alpha)} \ell^{2(1-2\al)/(1-\al)}
    U^{2\alpha}}{(2\alpha-1)} - R^2,
\end{equation}
where we have introduced the advanced and retarded coordinates
\begin{equation}\label{Advanced/Retarded}
    U = T + L, \quad V = T - L.
\end{equation}
This defines the $\Sigma_\ell$ hypersurfaces entirely in terms of
the $y^A$ coordinates and $\ell$.  This is a useful relationship
which we will need below.

To conclude, we have discovered that the $\Sigma_\ell$
hypersurfaces are constrained to lie in one half of
${\mathbb{M}}_5$ for all $\alpha > 0$.  Furthermore, they must
lie within the light cone centered at the origin for $\alpha \in
(\tfrac{1}{2},\infty)$ and outside it for $\alpha \in
(0,\tfrac{1}{2})$.  We have summarized all that we have learned
about the effects of $\alpha$ on the $\Sigma_\ell$ hypersurfaces
in Table \ref{table:summary}.
\begin{table*}
\begin{center}
\begin{tabular}{|c|c|c|c|c|} \hline
$\alpha$ interval & sign of $q(\tau)$ & osculating paraboloid &
caustic location & $T^2 - R^2 - L^2$
\\ \hline \hline $(0,\tfrac{1}{2})$ & $-$ & hyperbolic & null
infinity & $-$\\ $(\tfrac{1}{2},1)$ & $-$ & hyperbolic & origin &
$+$ \\ $(1,\infty)$ & $+$ & elliptical & origin & $+$ \\ \hline
\end{tabular}
\end{center}
\caption{Dynamical and topological properties of cosmologies
embedded on $\Sigma_\ell$ hypersurfaces. The ``caustic location''
refers to the point that the $\gamma_r$ lines approach as $t
\rightarrow 0^+$.}\label{table:summary}
\end{table*}

\subsection{Visualization of the $\Sigma_\ell$ hypersurfaces}
\label{sec:visualization}

\begin{figure}[t]
    \centering
    \includegraphics[height=2.5in]{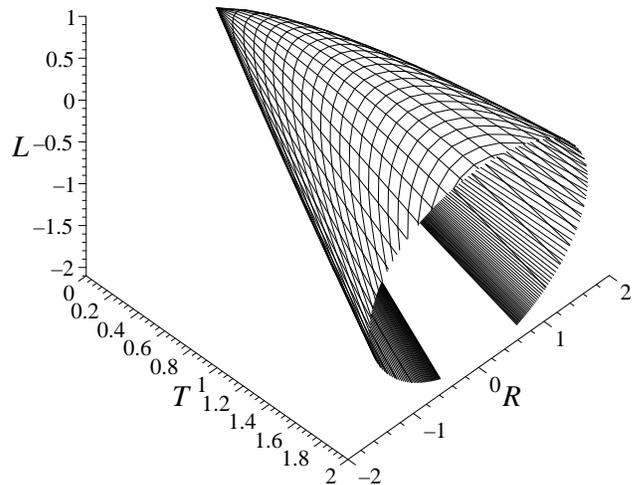}
    \caption{$\Sigma_\ell$ hypersurface with $\alpha = 3/2$ and
    $\ell_0 = 1$. We take $t\in[10^{-400},2]$ and $r\in[-10,10]$,
    with $\Delta t \approx 0.01$ and $\Delta r \approx 0.003$.}
    \label{fig:threehalves}
\end{figure}

\begin{figure}[t]
    \centering
    \includegraphics[height=2.5in]{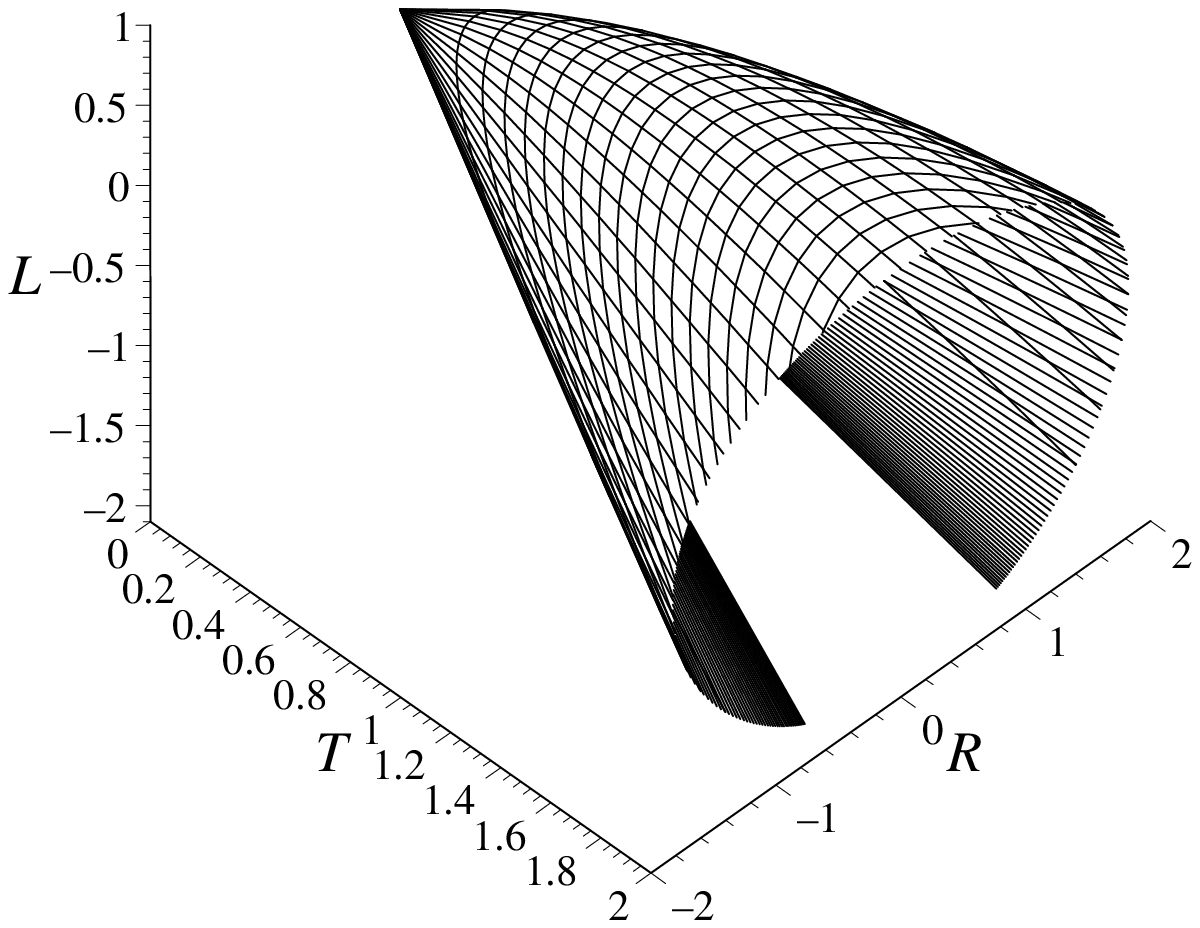}
    \caption{$\Sigma_\ell$ hypersurface with $\alpha = 2$ and $\ell_0 = 1$.
    We take $t\in[10^{-400},2]$ and $r\in[-10,10]$, with $\Delta t \approx 0.01$
    and $\Delta r \approx 0.003$.}
    \label{fig:two}
\end{figure}

\begin{figure}[t]
    \centering
    \includegraphics[height=2.5in]{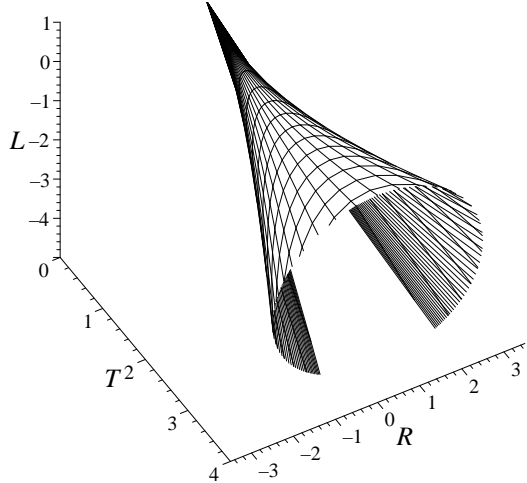}
    \caption{$\Sigma_\ell$ hypersurface with $\alpha = 2/3$ and $\ell_0 = 1$.
    We take $t\in[10^{-600},8]$ and $r\in[-3,3]$, with $\Delta t \approx 0.3$
    and $\Delta r \approx 0.08$.}
    \label{fig:twothirds}
\end{figure}

\begin{figure}[t]
    \centering
    \includegraphics[height=2.5in]{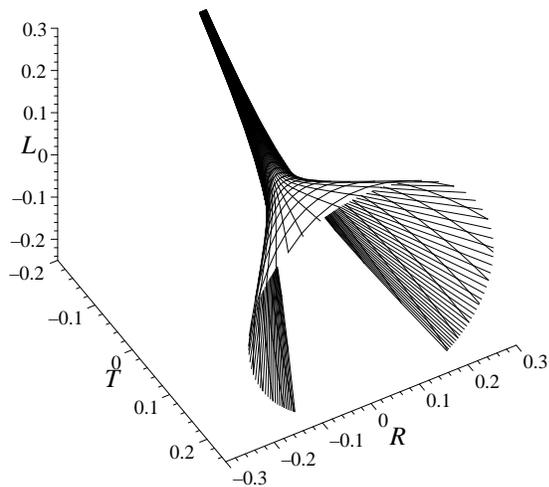}
    \caption{$\Sigma_\ell$ hypersurface with $\alpha = 1/30$ and $\ell_0 = 1$.
    We take $t\in[0.9,1.1]$ and $r\in[-0.1,0.1]$, with $\Delta t \approx 0.01$
    and $\Delta r \approx 0.003$.}
    \label{fig:oneoverthirty}
\end{figure}

\begin{figure}[t]
    \centering
    \includegraphics[height=2.5in]{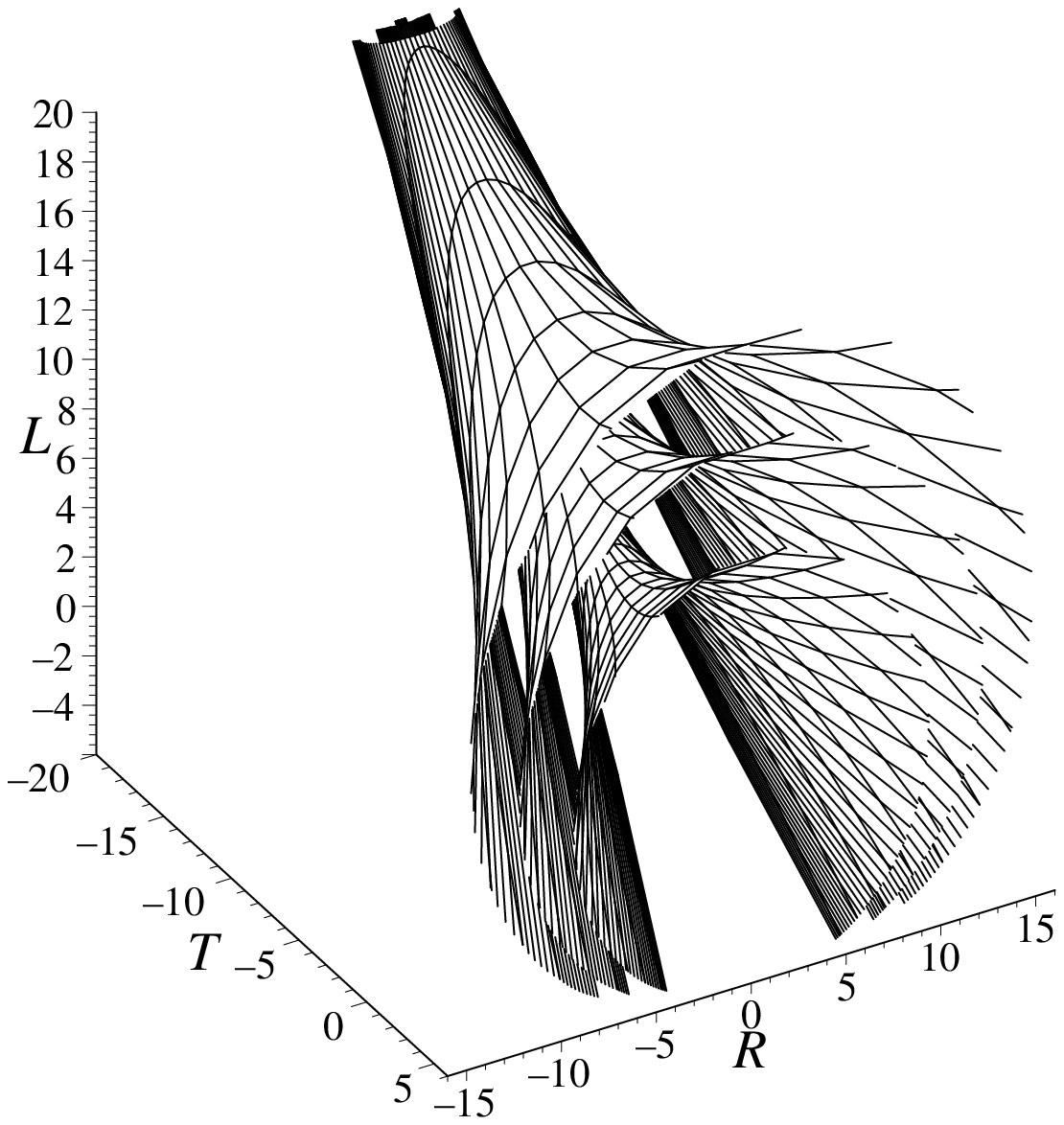}
    \caption{$\Sigma_\ell$ hypersurfaces with $\alpha = 1/3$ and
    $\ell_0 = 20,40,60$.  We take $t\in[10^{-300},3]$ and $r\in[-2,2]$,
    with $\Delta t \approx 0.05$ and $\Delta r \approx 0.07$.}\label{fig:many}
\end{figure}

We are now in a position to discuss Figures \ref{fig:threehalves}
-- \ref{fig:many} in detail.  The plots show the constant $t$ and
$r$ coordinate lines associated with a given $\Sigma_\ell$
hypersurface (denoted by $\gamma_t$ and $\gamma_r$ respectively)
as viewed from the $(T,R,L)$ space. As an aid to visualization,
we choose to let $r$ and $R$ range over both positive and
negative numbers, so the $\Sigma_\ell$ hypersurfaces are
symmetric about the $R=0$ plane (if it is desired to have $r>0$
or $R>0$ one of the symmetric halves can be deleted).  As the
models evolve in $t$-time they will generally grow in the
$R$-direction, so that the $\Sigma_\ell$ hypersurfaces appear to
``open-up'' in the direction of increasing $T$.  The $\gamma_t$
isochrones are seen to wrap around the hypersurfaces such that
they cross the $R=0$ plane perpendicularly.  The $\gamma_r$ lines
are orthogonal (in a Euclidean sense) to $\gamma_t$ lines at
$R=0$ and run along the length of the surface.

When using a computer to plot the surfaces, it is necessary to
choose finite ranges of $t$ and $r$.  The ranges used in each of
the Figures are given in the captions, along with the increments
in the time or radial coordinates between adjacent coordinate
lines (denoted by $\Delta t$ and $\Delta r$ respectively). For
models with $\alpha \in (\tfrac{1}{2},\infty)$, the $t=0$
``line'' is really a point for finite values of $r$.  Therefore,
the first $\gamma_t$ line in such plots has been chosen to be $t
= \delta$, where $\delta$ is the smallest positive number allowed
by machine precision.  For such models, the lower edge of the
$\Sigma_\ell$ hypersurfaces corresponds to the $t=\delta$ line.
For plots with $\alpha \in (0,\tfrac{1}{2})$ we have chosen to
restrict $t$ and $r$ to a fairly narrow ranges in order to
facilitate visualization. As a result of this, the lower edges of
$\Sigma_\ell$ in such plots are the $\gamma_r$ curves with the
largest value of $|r|$. However, we have confirmed that if the
range of $r$ is larger and $\min(t) = \delta$, then the comoving
trajectories with the largest values of $|r|$ tend to bunch up
along the $t = \delta$ line. That is, if the range of $r$ is
unrestricted then the lower edge of the $\Sigma_\ell$
hypersurfaces will correspond to the $\gamma_t$ line closest to
the big bang, irrespective of the value of $\alpha$.

We have carried out an extensive analysis of the morphology of
the $\Sigma_\ell$ hypersurfaces in the $(\alpha,\ell_0)$
parameter space, and present five informative cases which are
illustrated in the Figures:
\begin{description}
\item[Model 1:]
$\alpha = 3/2$, $\ell_0 = 1$.  This is the standard
matter-dominated model for the late universe (see above).  The
equation of state is that of dust and the scale factor evolves as
$t^{2/3}$. As expected, the osculating paraboloid has a
elliptical geometry for every non-singular point on
$\Sigma_\ell$.  The congruence of $\gamma_r$ curves has a caustic
at $T = R = L = 0$.  Notice that the rate of expansion of
$\gamma_r$ lines seems to slow as time progresses and that the
surface lies within the light cone centered at the origin.  This
plot is very similar to the embedding diagram shown in Figure 5
of Lynden-Bell et al.~\cite{Lyn89} (see the discussion in
subsection \ref{sec:line}).
\item[Model 2:]
$\alpha = 2$, $\ell_0 = 1$.  This is the standard
radiation-dominated model of the early universe (see above).  The
scale factor evolves as $t^{1/2}$, the caustic in the $\gamma_r$
congruence is located at the origin, and $\Sigma_\ell$ is
qualitatively similar to Model 1.
\item[Model 3:]
$\alpha = 2/3$, $\ell_0 = 1$.  A moderately inflationary model
with $(\rho+3p)<0$, a scale factor $\propto t^{3/2}$, and a
caustic in the $\gamma_r$ congruence at the origin.  The geometry
of the osculating paraboloid is hyperbolic, $\gamma_r$ lines
accelerate away from each other, and the entire surface lies
within the light cone centered at the origin.
\item[Model 4:]
$\alpha = 1/30$, $\ell_0 = 1$.  A very inflationary model with
$(\rho+3p)<0$, a scale factor $\propto t^{30}$, and the
$\gamma_r$ caustic at past null infinity. The osculating
paraboloid is hyperbolic with comoving paths flying apart. The
surface lies outside the light cone centered at the origin.
\item[Model 5:]
$\alpha = 1/3$, $\ell_0 = 20,40,60$.  A set of inflationary
models with $(\rho+3p) < 0$, a scale factor $\propto t^3$, and
the $\gamma_r$ caustic at past null infinity.  As expected, all
the surfaces lie outside the central light cone.  The space
between $\Sigma_\ell$ hypersurfaces and the null cone increases
with increasing $\ell$.
\end{description}
We note that the last case is based on classical field theory but
bears a striking resemblance to computer-generated models of
stochastic inflation \cite{Lin94} based on quantum field theory.

\subsection{Comments Regarding the Figures and the Geometrical Nature
of the Big Bang}
\label{sec:line}

One of the most interesting features of Figures
\ref{fig:threehalves} -- \ref{fig:many} is that the constant-$T$
cross sections $\gamma_T$ of the $\Sigma_\ell$ hypersurfaces are
very nearly closed. We will now demonstrate that the finite
amount of space between the lower edges of the surfaces depicted
in the Figures is a result of the requirement that $t \ge
\delta$, which is imposed by limitations in machine arithmetic.
If $r$ is eliminated from equations (\ref{T}) and (\ref{R}), we
find that the projection of $\gamma_t$ lines onto the $TR$-plane
is parabolic. For early times, we obtain
\begin{equation}
    T [2\alpha t^{1/\alpha} \ell^{1/(1-\alpha)} ] = R^2 -
    \frac{\alpha^2 t^2 \ell^2}{1-2\alpha}, \quad t \ll 1.
\end{equation}
If we hold $T$ constant and let $t \rightarrow 0^+$ then $R
\rightarrow 0$, which establishes that the $\gamma_T$ cross
sections are asymptotically closed.  Furthermore, since the lower
edges of the surfaces correspond to the minimum value of $t$, the
$\gamma_t$ isochrones must approach the $R = 0$ plane as $t
\rightarrow 0^+$.  By examining equations (\ref{light cone}) and
(\ref{holy grail}), it is easy to see that $\Sigma_\ell$ itself
must approach the $T+L =0$ plane as $t \rightarrow 0^+$.
Therefore, the $\gamma_t$ lines must approach some subset of the
$T+L= R = 0$ null line ${\mathcal{N}}$ as $t \rightarrow 0^+$,
which we denote by the intersection of ${\mathcal{N}}$ and
$\Sigma_\ell$: ${\mathcal{N}} \cap \Sigma_\ell$. In other words,
the finite gaps shown in the Figures are an artifact from
numerical calculations.

This raises a puzzling and subtle issue.  The $\gamma_t$ curves
approach a line in 5D at $t \rightarrow 0^+$, yet the coordinate
transformation (\ref{TL Transformation}) suggests that the $t=0$
``line'' is really a point in 5D for any finite value of $r$.  Is
the big bang a point-like or line-like event in 5D?  Several
authors have appreciated this problem in the past.  Lynden-Bell
et al.\ present an embedding diagram of a spatially-flat FRW
model with $p = 0$, which is equivalent to our Model 1 shown in
Figure \ref{fig:threehalves} \cite{Lyn89}.  Their plot is
qualitatively similar to ours, and they identify the half of
${\mathcal{N}}$ satisfying $T>0$ with the 4D big bang.  This is
contrary to the conclusion of Rindler, who demonstrates that open
FRW models may be completely foliated by spacelike 3-surfaces of
finite volume \cite{Rin00}.  The volume of these hypersurfaces
tends to zero as they approach the initial singularity, which
leads to the idea that the big bang is a point-like event.

We will now demonstrate that both of these views are in some
sense mathematically correct; the location of the big bang
depends on one's definition of the initial singularity.  One such
definition might be the locus of points defined by the positions
of the fundamental comoving observers in the $t \rightarrow 0^+$
limit. This is in the spirit of the definition of a singular
spacetime as a manifold containing one or more incomplete
timelike geodesics; i.e.\ timelike geodesics which are
inextendible in at least one direction, or have a finite affine
length \cite{Wal84}. The incomplete geodesics on $\Sigma_\ell$
are clearly the comoving trajectories of the $\gamma_r$
congruence, which has a fundamental caustic as $t \rightarrow
0^+$. For $\alpha \in (\tfrac{1}{2},\infty)$, it is obvious from
Figures \ref{fig:threehalves} -- \ref{fig:twothirds} that it is
impossible to extend these geodesics past the caustic at $T = R =
L = 0$.  The $\gamma_r$ geodesics are also incomplete for $\alpha
\in (0,\tfrac{1}{2})$, although this is hard to see from Figures
\ref{fig:oneoverthirty} and \ref{fig:many} because the caustic is
at null infinity.  However, we note that the proper time interval
along a comoving path between the initial caustic and any point
on $\Sigma_\ell$ is finite, which establishes that the $\gamma_r$
curves are incomplete.  In both cases, the comoving paths radiate
from a point in 5D. By identifying the location where the
$\gamma_r$ congruences ``begin'' with the initial singularity, we
conclude that the big bang is a point-like event in higher
dimensions.

However, the definition of the big bang need not be tied to the
properties of inextendible timelike geodesics.  We could instead
elect to define the location of the big bang as the locus of
points on $\Sigma_\ell$ where curvature scalars diverge.  This
definition is in the spirit of using quantities like the
Kretschmann scalar to distinguish between coordinate and genuine
singularities in the Schwarzschild spacetime (among others). For
the present case, we note that the Copernican principle built
into the FRW solutions demands that curvature scalars on
$\Sigma_\ell$ must be constant along $\gamma_t$ isochrones. This
is evidenced by the Kretschmann scalar (\ref{Kretschmann}) and
the Minkowski Gaussian curvature (\ref{Minkowski Gaussian}), both
of which demonstrate no dependence on $r$.  As $t \rightarrow
0^+$, both of these scalars diverge and the $\gamma_t$ isochrones
approach ${\mathcal{N}} \cap \Sigma_\ell$.  This would seem to
suggest the line ${\mathcal{N}} \cap \Sigma_\ell$ is the location
of a curvature scalar singularity on $\Sigma_\ell$.  By the
definition mentioned at the head of this paragraph, this implies
the big bang is a line-like event in higher dimensions.

There are two possible objections to this conclusion: Firstly,
the idea that $\Sigma_\ell$ is singular along $\mathrm{N} \cap
\Sigma_\ell$ is somewhat counterintuitive, because the
hypersurfaces shown in the Figures appear to smoothly approach
${\mathcal{N}}$ as $t \rightarrow 0^+$.  Secondly, we observe
that the $(t,r)$ coordinates do not actually cover ${\mathcal{N}}
\cap \Sigma_\ell$, which makes the $t$-dependent arguments of the
last paragraph suspect. This is easily seen by noting that the
induced metric (\ref{4D Metric}) is singular at $t = 0$ and $T+L
= t = 0$ on ${\mathcal{N}} \cap \Sigma_\ell$ from equations
(\ref{R}) and (\ref{T plus L}).  The situation is analogous to
the failure of spherical coordinates to cover the $z$-axis in
${\mathbb{E}}_3$. For these two reasons, it would be a good idea
to confirm the presence of a curvature singularity along
${\mathcal{N}} \cap \Sigma_\ell$ in a manner independent of the
$(t,r)$ coordinates.

This can be accomplished using the well-known extrinsic curvature
formalism.  Restoring the $\theta$ and $\phi$ coordinates, let us
now work in the 5D $\tilde{y}^A = (U,R,\theta,\phi,V)$ system
with $U$ and $V$ as defined in (\ref{Advanced/Retarded}). In this
coordinate system, the line ${\mathcal{N}} \cap \Sigma_\ell$ is
approached by taking the $U \rightarrow 0^+$ limit (recall that
$U \ge 0$ for all of the hypersurfaces).  We can define
$\Sigma_\ell$ in this space by
\begin{equation}
    0 = \Phi_\ell(U,R,V),
\end{equation}
where $\Phi_\ell(U,R,V)$ is given by the righthand side of
equation (\ref{holy grail}).  Here, $\ell$ is viewed a parameter
to be kept constant.  A vector normal to $\Sigma_\ell$ is then
given by
\begin{equation}
    N_A = \di_A \Phi_\ell(U,R,V).
\end{equation}
If $\Sigma_\ell$ is embedded in Minkowski space, then we find
that
\begin{equation}
    N^A N_A = -4 \ell^{(4\al-2)/(1-\al)}\al^{(2-4\al)}U^{2\al}
    \quad (\tilde{y}^A \in \Sigma_\ell \in {\mathbb{M}}_5),
\end{equation}
where we have used (\ref{holy grail}) to eliminate $R^2$, which
means that $N^A N_A$ is implicitly evaluated on $\Sigma_\ell$. We
note that $N^A$ is clearly null for $U = 0$, which corresponds to
${\mathcal{N}} \cap \Sigma_\ell$, and spacelike for $U > 0$,
which corresponds to the rest of $\Sigma_\ell$.  Hence, the
components of the unit normal to $\Sigma_\ell$, given by
\begin{equation}
    n^A = N^A/|N^B N_B|^{1/2},
\end{equation}
diverge as $U \rightarrow 0^+$.  Since the 5D manifold is flat,
the Riemann tensor on $\Sigma_\ell$ is related to the extrinsic
curvature 4-tensor in the following manner:
\begin{eqnarray}
    K_{\alpha\beta} & = & e^A_\alpha e^{B}_\beta \nabla_B n_{A}, \\
    R_{\alpha\beta\gamma\delta} & = & K_{\alpha\delta}K_{\beta\gamma}
    - K_{\alpha\gamma}K_{\beta\delta},
\end{eqnarray}
where $e^A_\alpha = \di \tilde{y}^A / \di \zeta^\alpha$ are
vectors spanning $\Sigma_\ell$ and $\zeta^\alpha$ are the
(arbitrary) coordinates used on $\Sigma_\ell$.  Because of the
bad behaviour of $n_A$ at $U = 0$, we fully expect the extrinsic
curvature and the 4D Riemann tensor to be singular along
${\mathcal{N}} \cap \Sigma_\ell$. Indeed, our extrinsic curvature
formalism is not really defined at $U=0$, so we must again be
content with a limiting argument. We can calculate various
curvature scalars composed from $K_{\alpha\beta}$ (or
equivalently $R_{\alpha\beta\gamma\delta}$) and show that they
diverge as $U \rightarrow 0^+$. For example, consider
\begin{equation}
    K^{\alpha\beta} K_{\alpha\beta} = (\nabla_B n^{A})(\nabla_A
    n^{B}),
\end{equation}
which can be established from the definition of $K_{\alpha\beta}$
and use of the induced metric $h_{AB} = h_{\alpha\beta}
e^\alpha_A e^\beta_B = g_{AB} + n_A n_B$.  For $\Sigma_\ell \in
{\mathbb{M}}_5$, we obtain via computer:
\begin{eqnarray}\nonumber
    K^{\alpha\beta} K_{\alpha\beta} & = & \frac{(\al^2 - 2\al +
    2)\al^{(2\al-2)} \ell^{(4\al-2)/(\al-1)}}{U^{2\al}} \\
    & \rightarrow & +\infty, \quad \text{as $U\rightarrow 0^+$
    ($\tilde{y}^A \in \Sigma_\ell \in {\mathbb{M}}_5$)}.
\end{eqnarray}
Therefore, $\Sigma_\ell$ has a curvature scalar singularity at
all points with $U = 0$; i.e.\ all along the line ${\mathcal{N}}
\cap \Sigma_\ell$.  This confirms the conclusion of the $t
\rightarrow 0^+$ limiting argument used above.

It is interesting to note that this curvature singularity is due
to $N^A$ becoming null along ${\mathcal{N}} \cap \Sigma_\ell$, as
opposed to any lack of ``smoothness'' of the surface at $U = 0$.
To see this, we can consider the embedding of $\Sigma_\ell$ in
Euclidean space.  Then, the norm of $N^A$ along ${\mathcal{N}}
\cap \Sigma_\ell$ (i.e.\ $U=0$) is
\begin{equation}
    N^A N_A = 2V^2, \quad [\tilde{y}^A \in ( {\mathcal{N}} \cap
    \Sigma_\ell ) \in {\mathbb{E}}_5].
\end{equation}
Unlike the Minkowski case, $N^A N_A$ only vanishes at a single
point on ${\mathcal{N}} \cap \Sigma_\ell$.  We also obtain via
computer
\begin{equation}
    K^{\alpha\beta} K_{\alpha\beta} =
    \begin{cases}
        0, & \alpha \in (0,\tfrac{1}{2}) \\
        2V^{-2}, & \alpha \in (\tfrac{1}{2},\infty)
    \end{cases},
\end{equation}
where $\tilde{y}^A \in ( {\mathcal{N}} \cap \Sigma_\ell ) \in
{\mathbb{E}}_5$.  For $\alpha \in (\tfrac{1}{2},\infty)$, we get
a curvature singularity for $U = V = R = 0$, which is the
position of the caustic in Figures \ref{fig:threehalves} --
\ref{fig:twothirds}. Interestingly enough, for $\alpha \in
(0,\tfrac{1}{2})$ we see no singularity at all, despite the fact
that $N^A N_A$ vanishes at $U = V = R = 0$.  This is because the
components of $N^A$ themselves vanish at $U = V = R = 0$, which
means that the unit normal $n^A$ is finite and well defined (as
may easily be verified). This is in agreement with Figures
\ref{fig:oneoverthirty} and \ref{fig:many}, which suggests that
$\Sigma_\ell$ is smooth at the origin.  We have also investigated
the behaviour of the full $K_{\alpha\beta}$ 4-tensor evaluated
along ${\mathcal{N}} \cap \Sigma_\ell$ in ${\mathbb{E}}_5$ and
confirmed that the components are in general well-behaved, except
at $V = 0$ for $\alpha \in (\tfrac{1}{2},\infty)$.  This
reinforces our conclusion that the only Euclidean curvature
singularity on $\Sigma_\ell$ is at $U = V = R = 0$ for $\alpha
\in (\tfrac{1}{2},\infty)$.  The chief difference between the
Minkowski and Euclidean embeddings is that $n^A$ is infinite
along ${\mathcal{N}} \cap \Sigma_\ell$ in the former case, while
in the latter scenario the unit normal is well defined all over
$\Sigma_\ell$.  Therefore, we can attribute the Minkowski
line-like curvature singularity on $\Sigma_\ell$ to the
divergence of the unit normal, and not to the lack of smoothness
of the surface along ${\mathcal{N}} \cap \Sigma_\ell$.

In conclusion, the geometric structure of the big bang in 5D
depends on one's definition of a singularity.  If a singular
location on $\Sigma_\ell$ is taken to be a place beyond which
timelike geodesics cannot be extended, then the big bang is a
point-like event in 5D.  If a singular location is taken to be a
place where curvature scalars on $\Sigma_\ell$ diverge, then the
big bang manifests itself as a line in 5D.  The distinction can
be elucidated by asking whether an observer at $r = 0$ at time $t
= t_0$ is ever in causal contact with points other than the
$\gamma_r$ caustic on ${\mathcal{N}} \cap \Sigma_\ell$.  The
trajectory of a light ray passing through $r=0$ at $t=t_0$ is
easily found from the null geodesic condition applied to the 4D
metric (\ref{4D Metric}):
\begin{equation}
    r(t) = \frac{\al \ell^{-\al/(1-\al)} [t_0^{(\al-1)/\al} -
    t^{(\al-1)/\al} ]} {(\al -1)}.
\end{equation}
Substituting for $r$ in (\ref{TL Transformation}) and taking the
$t \rightarrow 0^+$ limit, we find that light arriving at $r = 0$
at any finite time $t_0$ must have come from $T = R = L = 0$ for
$\al \in (\tfrac{1}{2},\infty)$ and $T = -L \rightarrow -\infty$,
$R = 0$ for $\al \in (0, \tfrac{1}{2})$.  In other words, the
only point on the ${\mathcal{N}} \cap \Sigma_\ell$ line inside
the cosmological horizon of $r = 0$ is coincident with the
position of the $\gamma_r$ caustic.  This is true for all
reasonable values of $\alpha$ and $t_0$.  Because of isotropy,
this must also be true for any finite value of $r$. Therefore, of
all the points on $\Sigma_\ell$ where curvature scalars diverge,
only the $\gamma_r$ caustic is in causal contact with ``normal''
points on $\Sigma_\ell$.  \emph{In a real physical sense, this
means that the geometric structure of the singularity that gives
rise to the observable universe is point-like in 5D.} Information
from the rest of the line-like singularity on $\Sigma_\ell$ can
never reach us within a finite amount of time. We feel that this
is the best possible resolution to the issue of the geometric
structure of the big bang in five dimensions.

\section{The $\alpha \rightarrow 0$ and $\alpha \rightarrow 1$
Limits, and the $\alpha = \tfrac{1}{2}$ Ponce de Leon Cosmology}
\label{sec:special cases}

The coordinate transformation (\ref{TL Transformation}) between
the original Ponce de Leon metric (\ref{5D Metric}) and the
Minkowski metric (\ref{5D Minkowski}) is undefined for $\alpha$ =
0, $\tfrac{1}{2}$, and 1. Of these three possibilities, only the
$\alpha = \tfrac{1}{2}$ case leads to a non-singular 5D metric
tensor (\ref{5D Metric}) in the $x^A$ coordinate system. We will
first discuss the cases for which the Ponce de Leon cosmologies
themselves are ill-defined and then turn to the highly-symmetric
$\alpha = \tfrac{1}{2}$ case.

In the limit $\alpha \rightarrow 0$, equation (\ref{Density and
Pressure}) implies that the equation of state of the induced
matter is that of de Sitter space, namely $\rho + p = 0$.  It is
well known that the scale factor of the de Sitter universe grows
exponentially in time, in contrast to the scale factor of the
Ponce de Leon cosmologies which can only grow as fast as a power
law. Therefore, it makes sense that de Sitter space corresponds
to the $\alpha \rightarrow 0$ limit of (\ref{5D Metric}) since it
is for this case that the scale factor $\propto t^{1/\alpha}$
grows faster than any (finite) power of $t$.  On the other hand,
the $\alpha \rightarrow 1$ case has an equation of state $\rho +
3p = 0$, which represents matter with no gravitational density by
equation (\ref{gravitational}). The limiting value of the scale
factor is $\propto t$ in this case, which is consistent with
$q(\tau) \rightarrow 0$ in equation (\ref{deceleration}).  This
limit clearly corresponds to the empty Milne universe, which is
known to have scale factor linear in time and matter with $\rho +
3p = 0$. It is interesting to note that $q(\tau) = 0$ would
correspond to the osculating paraboloid of $\Sigma_\ell$ being
cylindrical, which is precisely the situation intermediate to the
$\alpha = 3/2$ case shown in Figure \ref{fig:threehalves} and the
$\alpha = 2/3$ case shown in Figure \ref{fig:twothirds}.

Now, as mentioned above, the $\alpha = \tfrac{1}{2}$ cosmology is
the only case for which 5D metric (\ref{5D Metric}) is well
defined, but the transformation (\ref{TL Transformation}) is not.
This cosmology represents the case intermediate between universes
with a $\gamma_r$ caustic at $T=R=L=0$ (Figures
\ref{fig:threehalves} -- \ref{fig:twothirds}) and those with a
$\gamma_r$ caustic at null infinity (Figures
\ref{fig:oneoverthirty} and \ref{fig:many}). The line element by
(\ref{5D Metric}) is
\begin{equation}\label{onehalf}
    dS^2 = \ell^2 \, dt^2 - t^{4} \ell^4 \,
    d\sigma^2 - t^2 \, d\ell^2.
\end{equation}
By $t \rightarrow it$ this gives a metric with signature
$(----+)$, so the 4D spacetime is of the type used in Euclidean
quantum gravity \cite{Gib93}.  A further change $\ell \rightarrow
i\ell$ gives back (\ref{onehalf}).  The metric is also unchanged
for $t \leftrightarrow \ell$ with one or the other of $t
\rightarrow it$ and $\ell \rightarrow i\ell$.  However,
(\ref{onehalf}) with the last term taken with reversed sign is
not a solution of the field equations $R_{AB} = 0$, as may be
verified by computer.  It is therefore not like the so-called
``two-time'' 5D metrics with signature $(+---+)$ which satisfy
$R_{AB} = 0$ and have been applied to vacuum waves \cite{Bil96a}
and cosmological inhomogeneities \cite{Bil96b}.  It can also be
mentioned that the Ponce de Leon metric (\ref{5D Metric}) does
not satisfy $R_{AB} = 0$ if the last term is taken with reversed
sign.  Thus while (\ref{5D Metric}) and (\ref{onehalf}) satisfy
the field equations with a space-like extra dimension, they do
not with a time-like extra dimension, which raises the
possibility of testing the signature of higher-dimensional
theory.

The model (\ref{onehalf}) also shares with (\ref{5D Metric}) the
property that the 5D Riemann-Chistoffel tensor is $R_{ABCD} = 0$.
This may be verified by computer.  The scale factor in
(\ref{onehalf}) varies as $t^2$, which is faster than standard
FRW models.  It contains 4D matter which by (\ref{Density and
Pressure}) with $\tau \equiv \ell t$ is a perfect fluid with
density $\rho = 3/2\pi\tau^2$ and pressure $p=-1/\pi\tau^2$.  The
equation of state is therfore $p = -2\rho/3$.  The gravitational
density (\ref{gravitational}) is $(\rho + 3p) = -\rho$, so the
model is inflationary.

\section{Conclusion}\label{sec:conclusion}

$N$-dimensional extensions of general relativity and particle
physics are now numerous and include Kaluza-Klein theory,
induced-matter theory, superstrings, supergravity, string theory
and membrane theory.  Among these possibilities, the Ponce de
Leon models provide a concrete realization of cosmology in 5D
because they embed the FRW solutions of general relativity along
with their properties of matter.  Mathematically, a smooth local
embedding of 4D solutions with a finite energy-momentum tensor in
a class of 5D solutions of the vacuum field equations follows
from Campbell's theorem.  Physically, the 4D solutions exist on
hypersurfaces of the 5D manifold, and represent matter-filled
spacetimes because the 4D space is curved, even though in the
Ponce de Leon case the 5D space is flat.  What we have done in
the present account is to concentrate on the physical
consequences of how curved 4D cosmologies are embedded in the
flat 5D Ponce de Leon cosmologies.

One of our main results is to use the algebraic transformation
(\ref{TL Transformation}) from the models in FRW coordinates
(\ref{5D Metric}) to Minkowski coordinates (\ref{5D Minkowski})
to obtain pictures of the big bang.  Figures
\ref{fig:threehalves} -- \ref{fig:many} elucidate the structure
of several types of 4D universes by showing them as 2-surfaces
embedded in a 3D space.  The universes shown include standard
models for late (matter-dominated) and early
(radiation-dominated) epochs, as well as several inflationary
cases.  The congruence of comoving geodesics on any given
$\Sigma_\ell$ hypersurface is found to radiate from a caustic at
$t = 0$.  The caustic appears at different 5D positions for
models with $\alpha \in (0,\tfrac{1}{2})$ and $\alpha \in
(\tfrac{1}{2},\infty)$, but it always falls on the null ray
${\mathcal{N}}$.  The curvature of the hypersurfaces diverges
along the intersection of ${\mathcal{N}}$ and $\Sigma_\ell$, but
observers on $\Sigma_\ell$ are not in causal contact with any of
the singular points in ${\mathcal{N}} \cap \Sigma_\ell$ except
for the caustic in the comoving congruence.  This leads us to
conclude that the big bang is point-like in 5D.  The case
(\ref{onehalf}) where our coordinate transformation is undefined
has been investigated numerically and is also inflationary.

This and other aspects of our work would repay an in-depth study.
The main restriction on the results presented here is that they
refer only to the Ponce de Leon solutions of the vacuum 5D field
equations \cite{Pon88a}.  There is an extensive literature on
other solutions of the these field equations \cite{Wes99}. Some
of these are not 5D flat like the Ponce de Leon cosmologies.
However, we see no impediment to extending our approach using
embeddings to curved 5D solutions and $N(>5)$D solutions.

\begin{acknowledgments}
We thank A.D.\ Linde for discussions, NSERC for support and CIPA
for hospitality.  We would also like to thank the anonymous
referees, whose suggestions greatly improved the presentation of
our results.
\end{acknowledgments}


\begin{thebibliography}{10}

\bibitem{Haw73}
S.~W. Hawking and G.~F.~R. Ellis, {\em The Large Scale Strucutre
of Space-Time}
  (Cambridge University Press, Cambridge, 1973).

\bibitem{Wal84}
R.~M. Wald, {\em General Relativity} (University of Chicago
Press, Chicago,
  1984).

\bibitem{Lin90}
A.~D. Linde, {\em Inflation and Quantum Cosmology} (Academic
Press, Boston,
  1990).

\bibitem{Wes99}
P.~S. Wesson, {\em Space-Time-Matter} (World Scientific,
Singapore, 1999).

\bibitem{Lin79}
A.~D. Linde, Rep. Prog. Phys. {\bf 42},  389  (1979).

\bibitem{Bro79}
R. Brout, Phys. Rev. Lett. {\bf 43},  417  (1979).

\bibitem{Gut81}
A.~H. Guth, Phys. Rev. D {\bf 23},  347  (1981).

\bibitem{Hen83}
R.~N. Henriksen, A.~G. Emslie, and P.~S. Wesson, Phys. Rev. D
{\bf 27},  1219
  (1983).

\bibitem{Bon60}
W.~B. Bonner, J. Math. Mech. {\bf 9},  439  (1960).

\bibitem{Wes85}
P.~S. Wesson, Astron. Astrophys. {\bf 151},  276  (1985).

\bibitem{Gib93}
{\em Euclidean Quantum Gravity}, edited by G.~W. Gibbons and
S.~W. Hawking
  (World Scientific, Singapore, 1993).

\bibitem{Vil82}
A. Vilenkin, Phys. Lett. B {\bf 117},  25  (1982).

\bibitem{Dar00}
F. Darabi, W.~N. Sajko, and P.~S. Wesson, Class. Quant. Grav.
{\bf 17},  4357
  (2000).

\bibitem{Lid97}
J.~E. Lidsey, C. Romero, R. Tavakol, and S. Rippl, Class. Quant.
Grav. {\bf
  14},  865  (1997).

\bibitem{App87}
{\em Modern Kaluza-Klein Theories}, edited by T. Appelquist, A.
Chodos, and
  P.~G.~O. Freund (Addison-Wesley, Menlo-Park, 1987).

\bibitem{Cam26}
J.~E. Campbell, {\em A Course of Differential Geometry} (Claredon
Press,
  Oxford, 1926).

\bibitem{Rip95}
S. Rippl, C. Romero, and R. Tavakol, Class. Quant. Grav. {\bf
12},  2411
  (1995).

\bibitem{Rom96}
C. Romero, R. Tavakol, and R. Zalaletdinov, Gen. Rel. Grav. {\bf
28},  365
  (1996).

\bibitem{Wit81}
E. Witten, Nucl. Phys. B {\bf 186},  412  (1981).

\bibitem{West86}
P. West, {\em Introduction to Supersymmetry and Supergravity}
(World
  Scientific, Singapore, 1986).

\bibitem{Gre87}
M.~B. Green, J.~H. Schwarz, and E. Witten, {\em Superstring
Theory} (Cambridge
  University Press, Cambridge, 1987).

\bibitem{Duf96}
M.~J. Duff, Int. J. Mod. Phys. A {\bf 11},  5623  (1996).

\bibitem{You00}
D. Youm, hep-th/0004144  (2000).

\bibitem{Maa00}
R. Maartens, hep-th/0004166  (2000).

\bibitem{Cha00}
A. Chamblin, hep-th/0011128  (2000).

\bibitem{Lyn89}
D. Lynden-Bell, J. Katz, and I.~H. Redmount, Mon. Not. R. astr.
Soc. {\bf 239},
   201  (1989).

\bibitem{Rin00}
W. Rindler, Phys. Lett. A {\bf 276},  52  (2000).

\bibitem{Pon88a}
J. Ponce~de Leon, Gen. Rel. Grav. {\bf 20},  539  (1988).

\bibitem{Wes92}
P.~S. Wesson, Astrophys. J. {\bf 394},  19  (1992).

\bibitem{Wes94}
P.~S. Wesson, Astrophys. J. {\bf 436},  547  (1994).

\bibitem{Wes95}
P.~S. Wesson, Astrophys. J. {\bf 440},  1  (1995).

\bibitem{McM94}
D.~J. McManus, J. Math. Phys. {\bf 35},  4889  (1994).

\bibitem{Col95}
A.~A. Coley and D.~J. McManus, J. Math. Phys. {\bf 36},  335
(1995).

\bibitem{Abo96}
G. Abolghasem, A.~A. Coley, and D.~J. McManus, J. Math. Phys.
{\bf 37},  361
  (1996).

\bibitem{Lak95}
K. Lake, P.~J. Musgrave, and D. Pollney, {\em GR Tensor}
(Department of
  Physics, Queen's University, Kingston, Canada, 1995).

\bibitem{Liu95}
H. Liu and B. Mashhoon, Ann. Phys. (Leipzig) {\bf 4},  565
(1995).

\bibitem{Wes00}
P.~S. Wesson, H. Liu, and S.~S. Seahra, Astron. Astrophys. {\bf
358},  425 (2000).

%\bibitem{Ker00}
%R. Kerner, J. Martin, S. Mignemi, and J.-W. van Holten,
%gr-qc/0010098  (2000).

%\bibitem{Ell84}
%G.~F.~R. Ellis, Ann. Rev. Astron. Astrophys. {\bf 22},  157
%(1984).

%\bibitem{Pon88b}
%J. Ponce~de Leon, Phys. Rev. D {\bf 37},  309  (1988).

\bibitem{Lip69}
M. Lipshutz, {\em Theory and Problems of Differential Geometry},
{\em Schaum's
  Outline Series} (McGraw-Hill, New York, 1969).

\bibitem{Lin94}
A.~D. Linde, Scientific American {\bf 271},  48  (1994).

\bibitem{Bil96a}
A. Billyard and P.~S. Wesson, Gen. Rel. Grav. {\bf 28},  129
(1996).

\bibitem{Bil96b}
A. Billyard and P.~S. Wesson, Phys. Rev. D {\bf 53},  731
(1996).

%\bibitem{Dav86}
%A. Davidson and D.~A. Owen, Phys. Lett. B {\bf 177},  77  (1986).

%\bibitem{Sea01}
%S.~S. Seahra and P.~S. Wesson, in press in Gen. Rel. Grav.
%(2001).

%\bibitem{Hoy75}
%F. Hoyle, Astrophys. J. {\bf 196},  661  (1975).

%\bibitem{Wes88}
%P.~S. Wesson, Astron. Astrophys. {\bf 189},  4  (1988).

\end{thebibliography}
\end{document}